\documentclass[aps,pra,twocolumn,groupedaddress, showpacs]{revtex4}
 \usepackage{graphicx}
 \usepackage{amssymb}
 \usepackage{epstopdf}
\begin{document}
%
%  This is the real title on the first page.
\title{Feasibility of a storage ring for polar molecules in strong-field-seeking states}
\author{Hiroshi Nishimura}
\email{H_Nishimura@lbl.gov}
\affiliation{Mail Stop 80-101, Lawrence Berkeley National Laboratory, University of
California, Berkeley, California 94720}
\author{Glen Lambertson}
\email{GRLambertson@lbl.gov}
\author{Juris G. Kalnins}
\email{JGKalnins@lbl.gov}
\author{Harvey Gould}
\email{HAGould@lbl.gov}
\affiliation{Mail stop 71-259, Lawrence Berkeley National Laboratory,  Berkeley, CA 94720}

\date{\today}
\begin{abstract}
We show, through modeling and simulation,
that it is feasible to construct a storage ring
that will store dense bunches of strong-field-seeking polar molecules at 30 m/s
(kinetic energy of 2K) and hold them, for several minutes, against
losses due to defocusing, oscillations, and diffusion.
The ring, 3 m in diameter, has straight
sections that afford access to the stored molecules and a
lattice structure that may be adapted for evaporative cooling.
Simulation is done using a newly-developed code that
tracks the particles, in time, through 400
turns; it accounts for longitudinal velocity changes as a function
of external electric field, focusing and deflection nonlinearities,
and the effects of gravity. An injector, decelerator, and source are
included and intensities are calculated.
\end{abstract}
 \pacs{29.20 Dh, 41.75.Lx, 33.80.Ps, 39.90.+d, 33.55.Be}
 %end of abstract
%
\maketitle
\section{Introduction}
\label{intro} To date, all gaseous quantum condensates have been
produced by evaporative cooling of confined atoms. Confinement is
necessary to thermally isolate the particles from the warmer
environment and long confinement times are necessary because the
evaporative cooling process can take tens of seconds.

Strong magnetic field gradients have been used to confine neutral
paramagnetic molecules \cite{weinstein} and electric-field gradients
have been used to confine neutral polar molecules in electrostatic
traps \cite{bethlem00} and in toroidal storage rings
\cite{crompvoets01,crompvoets04}. In addition, polar molecule
confinement in a synchrotron storage ring has been modeled
\cite{nishimura03}.

All of these methods use molecules or atoms in
weak-field-seeking states, whose binding energy decreases in the field.
These states are not the lowest energy state and are therefore
subject to collisional relaxation.
In alkali atoms, the relaxation rates from the stretched hyperfine levels
($m_F = F$) is small. But in magnetically trapped
paramagnetic molecules \cite{volpi02} and in electrically confined
polar molecules \cite{bohn01,kajita01,kajita02,avdeenkov02}, the relaxation
rate can be large enough to
prevent achieving the confinement time needed for evaporative cooling.

Collisional relaxation will be absent for polar molecules in their
lowest rotational state. This ground state is strong-field-seeking,
as are all rotational states in the limit of strong electric field.
The technical challenges of storing molecules in a
strong-field-seeking state have not been previously addressed.
The major challenge is focusing these molecules because
electrostatic lenses can focus strong-field-seeking molecules in
only one transverse plane while defocusing in the other.
Therefore alternating-gradient focusing is required.

%For molecules in strong-field-seeking states, confinement in
%a storage ring is likely to be easier than in a trap.
%A storage ring can store many bunches of particles simultaneously,
%and with discrete focusing
%and bunching elements allows decoupling of the
%transverse and longitudinal motions, making beam cooling easier.
%The bunches in storage rings typically have
%longitudinal circulating velocities that are much higher
%than the velocity spread imposed by their temperature.

For experiments on molecules in strong-field-seeking states, a
storage ring has some useful features not generally found in traps.
The ring has a beam geometry with field-free regions accessible to
experiments, and it can simultaneously store many bunches of
particles producing a large flux of molecules.

% For one-column wide figures use
\begin{figure}
\begin{center}
% Use the relevant command for your figure-insertion program
% to insert the figure file.
% For example, with the option graphics use
\resizebox{0.45\textwidth}{!}{%
  \includegraphics{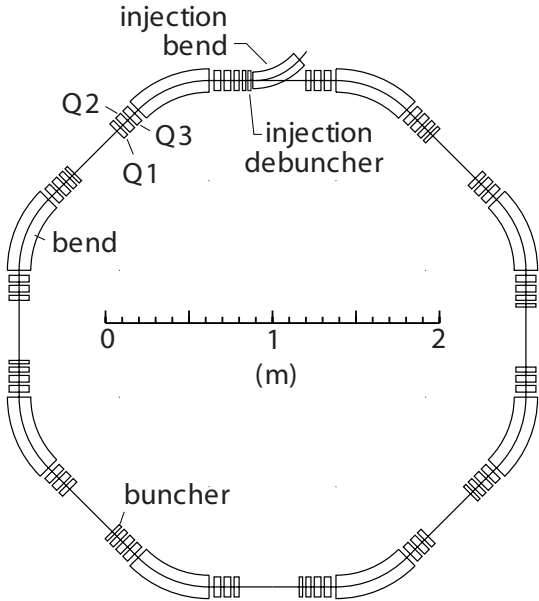}
} \caption{Layout of the storage ring. Each octant contains a buncher and
a pair of alternating-gradient focusing triplets to match the beam
traversing from the straight sections to the bend sections.
A bend section contains combined bend and
alternating gradient focusing elements.
The focusing and bend elements have time-independent electric fields.
An injection line is located in one of the straight sections}
\label{fig:RingLattice}       % Give a unique label
\end{center}
\end{figure}

In this paper we show, by modeling and simulation, that it is
feasible to construct a storage ring (Fig.\ref{fig:RingLattice})
that will store a symmetric-top molecule (methyl fluoride) in the
$J = 0$ state, at a kinetic energy of 2 K (30 m/s), and by extension
other molecules and velocities. In the storage ring, bunching
electrodes hold the molecules in a string of short bunches.
The molecules are calculated to be
stable against losses due to defocusing, oscillations, and diffusion
for over two minutes. We also model a decelerator for slowing the
molecules to 30 m/s, and an injector for loading the storage
ring.

A storage ring in which the density of the molecules
in a bunch is allowed to vary around the ring, can provide a
mechanism for evaporative cooling. Regions of high density speed
the thermalization of the molecules. In regions of low density
the molecules can become spatially separated due to their velocity spread
allowing the hottest molecules to be removed.

\section{Forces Due to Electric Field Gradients\label{sec:1}}

\subsection{Focusing and Deflection Using Multipole Fields}

A brief description of focusing and deflecting a beam of molecules
using electrostatic multipole fields is given below.
Additional details of beam transport and focusing of
molecules in strong-field-seeking states, with specific
application to methyl fluoride in the $J = 0$ state, may be found
in Kalnins et al. \cite{kalnins02}.

The guide field in a storage ring for molecules in
strong-field-seeking states must provide all the functions,
such as focusing,
bending, and bunching, that are used in a ring for charged
particles but with forces that arise from gradients of the
magnitude of the electric field.
%Those forces, in some similarity
%to the magnetic forces on a charged particle, if focusing in one
%transverse direction, can only defocus in the other.

In a pure quadrupole or sextupole field, the total electric field
increases radially and the force on a molecule, in a
strong-field-seeking state, is away from the centerline in all
transverse directions. Therefore a dipole component must be added to
remove the double-defocusing, and obtain focusing in one transverse
direction while still defocusing in the other. The force on a
molecule is given by the gradient of its Stark potential energy,
$W(E)$:
\begin{eqnarray}
   \emph{\textbf{F}}=-\nabla W(E)=-\frac{dW}{dE}\nabla E\label{eq:Force}
\end{eqnarray}
where $E$ is the magnitude of an external field.

The Stark energy of the molecular level is in general a nonlinear
function and is described for methyl fluoride in the $J = 0$
rotational state in Ref. \cite{kalnins02}. In the limit of large $E$,
$W(E)\rightarrow-d_eE$ where $d_e$ is the molecule's electric dipole
moment.

The transverse ($x$ horizontal, $y$ vertical) electric
multipole potential used to bend and focus a molecule is:
\begin{eqnarray}
  \Psi=E_0[y+A_2xy+A_3(x^2y-\frac{1}{3}y^3)] \label{eq:PotPsi}
\end{eqnarray}
where $E_0$ is the dipole field strength, and $A_2$ and $A_3$ are
the relative quadrupole and sextupole component strengths.

For the Stark energy in the high-field limit, the forces to second
order are:
\begin{eqnarray}
        F_x&\rightarrow&d_eE_0[A_2+2A_3x-\frac{1}{2}(A_2^3-4A_2A_3)y^2]\nonumber \\
        F_y&\rightarrow&d_eE_0[(A_2^2-2A_3)y-(A_2^3-4A_2A_3)xy ]\label{eq:PhiXYQ}
\end{eqnarray}

We see that a combined dipole and sextupole ($A_3$) field
lens will focus in one plane, while defocusing in the other.
To deflect the molecule we must add a quadrupole ($A_2$) component.
This also defocuses the beam in the $y$ direction and stronger
sextupole ($A_3$) strengths are needed \cite{OurPAC2003paper}.

To obtain net focusing in both transverse planes, the lenses are
arranged in a sequence with gradients alternating in sign ($A_3 <0$
for $x$-focusing and $A_3>$0 for $y$-focusing).

%\begin{eqnarray}
%  \Psi=E_0[y+A_2xy+A_3(x^2y-\frac{1}{3}y^3)] \label{eq:PotPsi}
%\end{eqnarray}
%
%where we use x for horizontal and and y for vertical coordinate. The
%forces in the direction $x$ and $y$ respectively are:
%\begin{eqnarray}
%        F_x&\rightarrow&d_eE_0[A_2+2A_3x-\frac{1}{2}(A_2^3-4A_2A_3)y^2\nonumber \\
%           & &\phantom{d_eE_0}+(A_2^4-5A_2^2A_3^2+4A_3^2)xy^2]\nonumber \\
%        F_y&\rightarrow&d_eE_0[(A_2^2-2A_3)y-(A_2^3-4A_2A_3)xy\\
%           & &\phantom{d_eE_0}+(A_2^4-5A_2^2A_3^2+4A_3^2)x^2y]\nonumber
%\label{eq:PhiXYQ}
%\end{eqnarray}

\subsection{Other Effects}

When a molecule in a strong-field-seeking state
enters the field of an electrode pair
it is accelerated longitudinally, and upon exiting the field it is decelerated.
Also, the
fringing field is stronger away from the midplane and this causes
a net defocusing force in the direction of the electric field. Between
successive sets of electrodes, this unwanted defocusing is reduced
if the dipole fields are of the same polarity and strength.

Longitudinal bunching, as in a charged-particle ring, requires a
pulsed field. The field is ramped in a sawtooth or sine-wave form
and the time-dependent acceleration is the net difference between
the fields when entering and when exiting.

The effect of gravity is small but not negligible
for 30 m/s molecules in this ring. The vertical orbit will be
distorted and an orbit correction must be applied.

\subsection{Equations of Motion}

%The equations of motion of a molecule in the ring are obtained from
%the Hamiltonian  $H=H_0 + W + U$, where U is the gravitational
%potential and W(E) is the Stark potential of methyl fluoride in the
%guide field of the ring.

The equations of motion of a molecule in the ring are obtained
from the Hamiltonian:
\begin{eqnarray}
H = H_0 +W(E) - gy
\end{eqnarray}
where $W(E)$ is the Stark energy, g is the acceleration due to gravity,
and $H_0$ is the kinetic energy which in a bend region is:
\begin{eqnarray}
H_{0}=\frac{1}{2m}(P_{x}^2+P_{y}^2+\frac{P_\theta^2}{(\rho+x)^{2}})\label{eq:H0bend2}
\end{eqnarray}
where $P_x$ and $P_y$ are the transverse momenta,
$P_{\theta}$ is the angular momentum and $\rho$ is the
bend radius. In straight sections the last term is replaced by
the square of the longitudinal momentum, $P{_z}^2$.

The longitudinal variation of the Stark energy at the ends of
electrodes (treated here as a step function) adds or subtracts from the
kinetic energy, the change in longitudinal velocity being about
$\pm$10$\%$.

Vertical defocusing in a fringe field is derived from the
longitudinal variation of the field on the midplane and to lowest
order is:
\begin{eqnarray}
%% m\ddot{y}=-\frac{dW}{dE}y
 (F_y)_{fringe}=-\frac{dW}{dE}\phantom{a}
   [\frac{1}{E_y}( \frac{\partial E_y}{\partial z })^2
    -\frac{\partial ^2 E_y}{\partial z^2}]\phantom{a}y \label{eq:Edge}
\end{eqnarray}

\section{Storage Ring Design}
\label{sec:4}
\subsection{Molecule and Energy}

The principles and techniques we use apply to all polar molecules in
strong-field-seeking states. We choose methyl fluoride (CH$_3$F) as
our reference molecule because it is a nearly symmetric rotor with a
large electric dipole moment of $d_e$ = $6.2 \times 10^{-30}$ C-m
(1.84 D).
It has a moderate rotational
constant of $B = 0.88$ cm$^{-1}$ and a simple level structure
with a $J = K = 0$
rotational ground state. The rotational constant is large enough to
limit the number of rotational levels populated in the beam from a
jet-source but still small enough to allow for a large Stark effect
at moderate electric fields. Methyl fluoride is also a gas at room
temperature.

The velocity of 30 m/s (kinetic energy of about 2K) is low enough to
make for a compact ring, yet keep small the effects of gravity.

\subsection{Ring Lattice}
\label{ring}

 Long straight regions free of focusing electrodes make the stored
beam accessible for experiments and give space for injection and
extraction.  Molecules, in order to drift through the straight
section without loss, must have only small divergences and
therefore a large beam width.
In a bending region, we need strong deflecting forces
to minimize the bend radius for overall compactness.
These strong forces call for a small beam width to avoid
nonlinearities.  To make the transition (match) from straight sections to
arc sections, triplets (Q1, Q2, Q3) of focusing lenses
are placed at the ends of the
straight sections, as shown in Fig. \ref{fig:RingLattice}.

In each of the eight bend regions, there are five electrode pairs;
each has a
combined dipole and quadrupole field to provide the strong
deflecting force. To this is added a sextupole component, the
gradient of which alternates in sign.

The electrode parameters are given in Table
 \ref{Table:QuadParam} where Q are focusing elements
 and BF and BD are combined bend and focusing elements.
 Each arc is a series of BF and BD elements:
  $\frac{1}{2}$BF+BD+BF+BD+$\frac{1}{2}$BF.

\begin{figure}
\begin{center}
% Use the relevant command for your figure-insertion program
% to insert the figure file.
% For example, with the option graphics use
\resizebox{0.35\textwidth}{!}{%
  \includegraphics{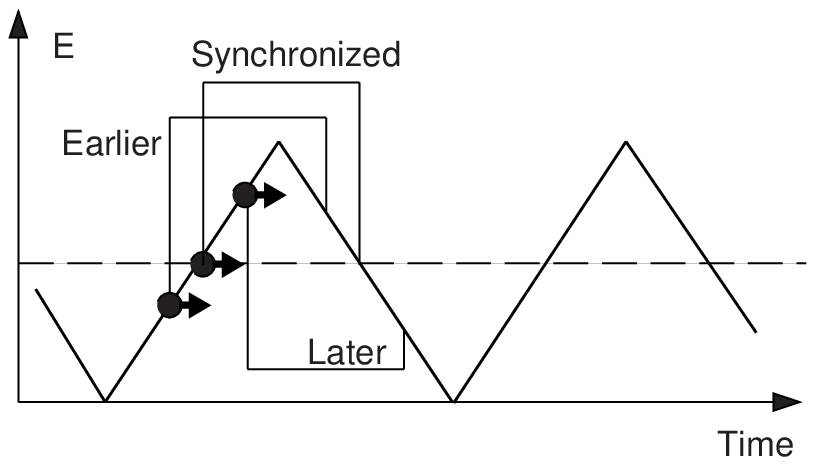}
} \caption{A molecule at the bunch center
enters and exits the buncher when the field is the same
and receives no net acceleration. For a molecule that arrives later,
the entering
field is stronger than at its exit; it is accelerated and it then
drifts downstream toward the bunch center.}
\label{fig:dect}       % Give a unique label
\end{center}
\end{figure}

In this sequence of lenses with alternating gradients, the molecules
execute oscillatory transverse motions.  The parameters of BF and BD
are chosen such that the phases of these horizontal and vertical
motions each advance through an angle of 2$\pi$ in each octant of
arc. The parameters of Q1, Q2 and Q3 are varied to
 find values that produce large
 dynamic aperture and momentum acceptance.
 The decapole
 coefficient, $A_5$ of Q2, which adds the term
 $E_0A_5(x^4y-2x^2y^2+\frac{1}{5}y^5)$ to the potential
 Eq(\ref{eq:PotPsi}), is introduced to reduce the nonlinearity
 of Q2 focusing where the beam is at it's largest.
 For longitudinal confinement with many short bunches,
 we use eight bunchers in the ring; each has
a short uniform field that is pulsed in time as illustrated in Fig.
\ref{fig:dect}.

Molecules with different energies have their closed orbits radially
separated in the arcs and perhaps elsewhere in the ring. If this
dispersion of orbits is present at a buncher, the energy change from
the buncher produces a shift in the orbit and an increment in the
radial oscillation. This is called synchro-betatron coupling and to
avoid growth of radial oscillation amplitude, the dispersion of
orbits must be made zero at the bunchers. With the phases of the
vertical and horizontal motions advancing through an angle of $2\pi$
 in each octant, as noted above, the
dispersion becomes zero at all eight buncher locations.

%With the phase advance of
%$2\pi$ in each octant, noted above, the dispersion becomes zero at
%all eight buncher locations.

\begin{table}[htbp]
\begin{center}
\caption{Parameters of Storage Ring Electrodes}
\begin{tabular}{|l|c|c|c|c|c|}
\hline
      &$Eo$&$L$&$A_2$&$A_3$&$A_5$\\
      &(MV/m)&(cm)&(m$^{-2}$)&(m$^{-3}$)&(m$^{-5}$)\\
\hline
        Q1   & 3.0  & 3.34 & 0    & 2000&       0\\
        Q2   & 4.0  & 3.71 & 0    &-2000&-1.28$\times$10$^6$\\
        Q3   & 4.0  & 2.85 & 0    & 2000&       0\\
        BF   & 7.85 & 4.00 &-10.55&-2296&0\\
        BD   & 7.85 & 4.00 &-10.55& 2343&0\\
\hline
\end{tabular}
\label{Table:QuadParam}
\end{center}
\end{table}

\subsection{Numerical Modeling and Simulation}

\begin{figure}[htbp]
\begin{center}
\resizebox{0.50\textwidth}{!}{%
  \includegraphics{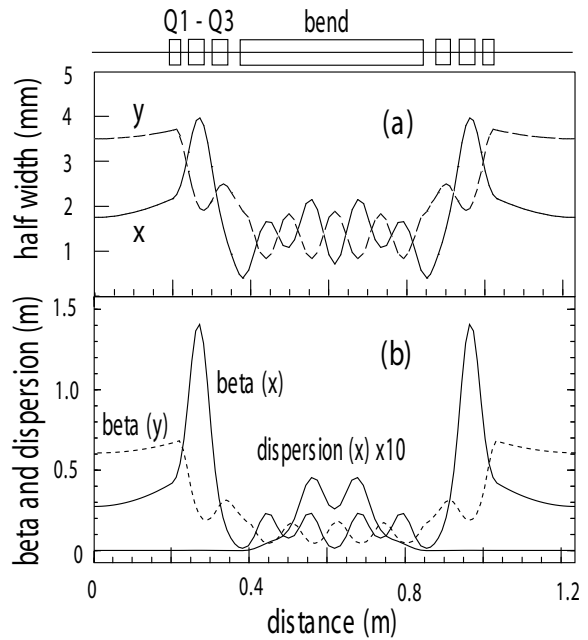}
}
    \caption{Beam half-widths (a) and the beta functions and
    horizontal dispersion (b) in
    the storage ring.
    Beta is the distance in which the transverse
(betatron) oscillation advances in phase by one radian. A schematic of the
lattice is shown for location reference.}
    \label{fig.TwissCH3F}
\end{center}
\end{figure}

\begin{figure}[htbp]
\begin{center}
% Use the relevant command for your figure-insertion program
% to insert the figure file.
% For example, with the option graphics use
\resizebox{0.35\textwidth}{!}{%
  \includegraphics{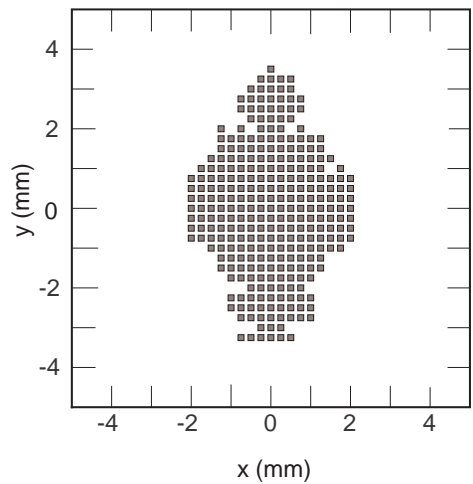}
}
%\caption{Dynamic Aperture of the CH3F Ring}
\caption{Starting coordinates in the center of the straight
section for the molecules that survive 400 turns.
This defines the dynamic aperture.}
\label{fig:DynapCH3F}       % Give a unique label
\end{center}
\end{figure}

The lattice parameters (Table \ref{Table:QuadParam})
are found by
numerical calculations using a newly-developed simulation code that
tracks the particles in time (rather than in longitudinal position)
to account for the longitudinal velocity changes as a function of
the external field. The tracking code includes the effects of
nonlinearities, gravity and the longitudinal kick at the bunchers.
The effect of each fringe field (Eq. \ref{eq:Edge}) in every element
has been integrated and replaced by a vertically defocusing thin lens.
The parameters in Table \ref{Table:QuadParam} result in the
ring performance listed in Table.
\ref{Table:MainParam} and shown in Figures \ref{fig.TwissCH3F} and
\ref{fig:DynapCH3F}.

\begin{table}[hbt]
\begin{center}
\caption{Ring Parameters}
\begin{tabular}{|l|c|}
\hline
 Parameter&Value\\
\hline
           Circumference (m)                & 9.850 \\
           Circulation period (s)           & 0.3121\\
           Velocity in free space (m/s)     & 30.0\\
           Symmetry of the ring             & 8  \\
           Bending radius (m)               & 0.60 \\
           Long straight section (m)        & 0.40 \\
           Beta function$^*$ $\beta_x$ (m)& 0.274 \\
           \phantom{Beta function$^*$ }$\beta_y$ (m)& 0.596 \\
           Dispersion$^*$ $\eta_x$ (m)& 0.0\\
           Betatron tune $\nu_x$ & 13.368 \\
           \phantom{Betatron tune }$\nu_y$   &   10.398 \\
           Dynamic aperture$^*$ $a_x$ (mm) & $\pm$1.75\\
           \phantom{Dynamic aperture$^*$ }$a_y$ (mm) & $\pm$3.50\\
           Acceptance $\epsilon_x$ (mm - mr) & 11\\
           \phantom{Acceptance }$\epsilon_y$ (mm - mr) & 21\\
           Momentum acceptance ($\%$) & $\pm$1.2\\
           Number of longitudinal buckets  & 203 \\
\hline
\end{tabular}
\begin{tabular}{l}
$^*$At the center of straight sections
\end{tabular}
\label{Table:MainParam}
\end{center}
\end{table}

The beta functions and the horizontal dispersion are shown in  Fig.
 \ref{fig.TwissCH3F}b.
 Small beta functions in the bends produce a smaller beam profile,
 allowing the bend elements to be stronger and the beam to occupy
 the most linear region of the elements.
The straight sections are designed to be free of horizontal dispersion
to avoid synchro-beta coupling at the bunchers.

\begin{figure}[htbp]
\begin{center}
% Use the relevant command for your figure-insertion program
% to insert the figure file.
% For example, with the option graphics use
\resizebox{0.40\textwidth}{!}{%
  \includegraphics{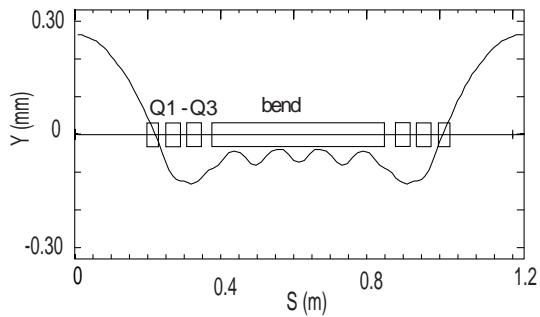}
} \caption{Corrected vertical closed orbit displacement
of the beam in the storage ring}
\label{fig:CODCH3FWC}       % Give a unique label
\end{center}
\end{figure}

If uncorrected, the vertical closed orbit displacement caused
by gravity is 2.6 mm
and is large enough to cause loss of the circulating beam.
The orbit is corrected by displacing Q2 by 0.24 mm downward to
produce upward kicks. The resulting vertical orbit
distortion shrinks to 0.26 mm as shown in Fig. \ref{fig:CODCH3FWC} and is
no longer a problem.

With this orbit correction, the dynamic aperture for 400 turns,
at the center of a straight section, is about 2 mm by 3 mm
half-width as shown in Fig. \ref{fig:DynapCH3F}.
 This dynamic aperture corresponds to acceptances
of 11 mm-mr horizontal and 21 mm-mr vertical, as listed in Table
\ref{Table:MainParam}. The resulting beam size is shown in Fig.
\ref{fig.TwissCH3F}a. The momentum acceptance, calculated by the
multi-particle tracking simulation, is $\pm 1.2\%$ which is
equivalent to an energy acceptance of $\pm$ 45 mK.

\section{Decelerated Beam}
\label{sec:6}
\subsection{Decelerator}

To reduce the velocity from the 310 m/s at the source
to 30 m/s requires many
stages of deceleration by pulsed electric fields in a long linear
array. At each of the 139 decelerating stages, a bunch of molecules
enters a set of parallel electrodes when the field is zero;
the field pulses on and the
molecules lose kinetic energy equal to $|W(E)|$ as they exit the electrodes.

Our decelerator design differs in almost every way from previous
designs \cite{bethlem02a,tarbutt04}. A decrease in the strength of
the electric field while the bunch exits the electrodes provides
longitudinal restoring action that prevents the bunch lengthening
due to velocity spread \cite{maddi99}. The lengths of successive
electrodes decreases as the velocity and spacing of the bunches
decreases.

Interspersed between the pulsed parallel electrodes
are alternating-gradient lenses to confine the
molecules transversely. Their overall focusing action must be
stronger in the plane of the electric fields to counter the
defocusing from fringe fields.
The major parameters of the decelerator are summarized in
Table \ref{table:ParamDecelInjection}.
Details of decelerator design will be published later.

% For tables use
\begin{table}
\caption{Parameters of the decelerator for injected beam}
\label{table:ParamDecelInjection}
\begin{center}
\begin{tabular}{|l|c|}
\hline
 Parameter&Value\\
\hline
    Velocity at source (m/s)&310\\
    Velocity at exit (m/s)&30\\
    Velocity spread at exit (\%)&$\pm$2\\
    Length of bunch at exit (mm)&10\\
    Emittances at exit, x and y (mm-mr)&30\\
    Electrode gap (mm)&7\\
    Decelerating field at entrance (MV/m)&9\\
    Decelerating field at exit (MV/m)&4.5\\
    Length of last decel. electrode (mm)&24\\
    Length of decelerator (m)&19.6\\
    Number of decel. electrodes&139\\
\hline
\end{tabular}
\end{center}
\end{table}

\subsection{Injector}
To inject the beam, we need a bend electrode that can pulse on or
off in the time between buckets in the ring.
This allows us to store multiple (up to 203) bunches in the ring.
The deflecting electrode (Fig.\ref{fig:RingLattice}) is part of a
transport line that transforms the pulse leaving the decelerator to match the
orientation of the transverse acceptances of the ring at the point
of entry onto the closed orbit of the ring.  The deflecting
electrode is actually an array of bend electrodes with radius 0.6 m,
similar to a bend section in the ring. A horizontal phase advance of
$2\pi$ in this bend, avoids a net dispersal of molecules that are within the
$\pm 2$ \% velocity spread.

In passage along the line, the velocity
spread of $\pm 2$\% lengthens the bunch and a debuncher at the point
of injection (Fig \ref{fig:RingLattice})
brings 90\% of the bunch within the $\pm 1.2$\% longitudinal momentum
acceptance of the ring.

%\begin{figure}[htbp]
%\resizebox{0.40\textwidth}{!}{%
%  \includegraphics{UnpreparedFigure3}
%} \caption{Injection line} \label{fig:injection-line}
%\end{figure}

\subsection{Source and Intensity}
We calculate the intensity based upon a pulsed jet source with 1\% methyl
fluoride seeded in xenon carrier gas, using the equations in Miller
 \cite{miller88} and verified against seeded xenon jet source
performance reported in the literature
\cite{crompvoets01,crompvoets04,gupta99}. Xenon's high mass (133)
produces much slower beams (310 m/s from a room-temperature
reservoir) than do light carrier gases, resulting in a shorter (19.6
m) decelerator.

The bunch intensity is determined by the source flow rate, the $J = 0$ state
population, the velocity distribution and the acceptances.
A source orifice of 1 mm diameter and reservoir pressure
of $6.56 \times 10^4$ Pa (500 Torr) %will produce an intense cold beam of
will produce an intense cold beam with a peak intensity of $3 \times
10^{19}$ molecules sr$^{-1}$ s$^{-1}$, a longitudinal velocity
spread of 7.2 m/s FWHM, and less than 1\% clusters. We estimate the
methyl fluoride J = 0 rotational state fraction to be 30\%. In an
apparatus with a finite pumping speed, this peak intensity is
possibly only by using a pulsed jet source operating with a small duty
cycle. The short widely-spaced beam pulses entering the decelerator
(which become more closely spaced after deceleration) require a 
duty cycle of less than one percent for a 100 Hz pulse rate.
This would allow all 203 buckets in the ring to be filled in 6.4
turns.

The transverse and the longitudinal emittances (units of m$^2$ s$^{-1}$) of a
bunch of molecules are unchanged in passing through the deceleration
process\cite{lambertson04} from the source to their injection in the storage ring.  
Therefore the
fraction of molecules from the source that enters the ring is the
product of the ratios of ring acceptances to source emittances. 
In the transverse directions, the beam from the source has $\pm 0.5$ mm spatial
extents and $\pm 1000$ mr angular divergences; then the horizontal and
vertical acceptances of the storage ring (Table \ref{Table:MainParam}) 
of 11 mm-mr and 21
mm-mr respectively, result in $8.66 \times 10^{-6}$ of the molecules being
transversely accepted.  Longitudinally, one second of beam from the
source is 310 m long and has a velocity spread of $\pm 3.6$ m/s. The storage
ring will accept $\pm 0.6$ m/s in a 10-mm long bunch, which is 
$5.4 \times 10^{-6}$ of the source longitudinal emittance.

Combining these nunbers and acounting for the 90\% acceptance of the
storage ring from the injector yields an intensity of $3.8 \times
10^8$ molecules/bunch. Bunches could be injected into the storage
ring singularly or in large numbers. With a maximum of 203 stored
bunches there would be nearly 10$^{11}$ molecules
circulating in the storage ring and a flux of 2.5 $\times$ 10$^{11}$
molecules/s. Each bunch would have a density of about 3 $\times
$10$^9$ molecules/cm$^3$ in the long straight sections, and higher
in the bends.

\section{Acknowledgments}
The authors acknowledge and thank Richard Gough and
David Robin for their enthusiastic encouragement,
and Swapan Chattopadhyay and Ying Wu
for early contributions to the storage ring work.
Work supported by the Director,
Office of Science; Office of Basic Energy Sciences, and
Office of High Energy and Nuclear Physics, U.S. Department of
Energy, under Contract No. DE-AC03-76SF00098.


\begin{thebibliography}{}
%
% and use \bibitem to create references.
%
\bibitem{weinstein}
J. D. Weinstein, R. deCarvalho, T. Guillet, B. Friedrich
and J.M. Doyle, Nature \textbf{395}, 148 (1998).
%
\bibitem{bethlem00}
H.L. Bethlem, G. Berden, F.M.H. Crompvoets, R.T. Jongma,
A.J.A. van Roij, and G. Meijer, Nature \textbf{406}, 491 (2000).

\bibitem{crompvoets01}
F.M.H. Crompvoets, H.L. Bethlem, R.T. Jongma, and G. Meijer,
Nature \textbf{411}, 174 (2001).

\bibitem{crompvoets04}
F.M.H. Crompvoets, H.L. Bethlem, J. K\"{u}pper, A.J.A. van Roij and
G. Meijer, Phys. Rev. A\textbf{69}, 063406 (2001).

\bibitem{nishimura03}
H. Nishimura, G. Lambertson, J. G. Kalnins, and H. Gould, Rev. Sci.
Instr. \textbf{43}, 3271 (2003).

\bibitem{volpi02}
A. Volpi and J.L. Bohn, Phys.\ Rev.\ A\textbf{65}, 052712 (2002).

\bibitem{bohn01}
J. L. Bohn, Phys.\ Rev.\
A\textbf{63}, 052714 (2001).

\bibitem{kajita01}
M. Kajita, T. Suzuki, H. Odashima, Y. Moriwaki, and M. Tachikawa,
Jpn. J. Appl. Phys. \textbf{40}, L1260 (2001).

\bibitem{kajita02}
M. Kajita, Eur. Phys. J. D\textbf{20}, 55 (2002).

\bibitem{avdeenkov02}
A.V Avdeenkov and J.L. Bohn,
Phys.\ Rev.\ A\textbf{66}, 0052718 (2002).

\bibitem{kalnins02}
J.G. Kalnins, G. Lambertson, and H. Gould, Rev. Sci. Instr.
\textbf{73}, 2557 (2002).

\bibitem{OurPAC2003paper}
H. Nishimura, G. Lambertson, J.G. Kalnins, and H. Gould, IEEE Proc.
of PAC 2003, 1837 (2002).

\bibitem{bethlem02a}
H.L. Bethlem, A.J.A. van Roij, R.T. Jongma and G. Meijer, Phys. Rev.
Lett. \textbf{88}, 133003 (2002).

\bibitem{tarbutt04}
M.R. Tarbutt, H.L. Bethlem, J.J. Hudson, V.L. Ryabov, V.A. Ryzhov,
B.E. Sauer, G. Meijer, and E.A. Hinds, Phys. Rev. Lett. \textbf{92},
173002 (2004).

\bibitem{maddi99}
J.A. Maddi, T.P. Dinneen, and H. Gould, Phys.\ Rev.\, A\textbf{60},
3882 (1999).

\bibitem{miller88}
D.R. Miller, in \emph{Atomic and Molecular Beam Methods}, edited by
G. Scoles (Oxford University Press, New York, 1988), Vol. 1, p. 14.

\bibitem{gupta99} M. Gupta and D. Herschbach,
J. Phys. Chem. \textbf{103}, 10670 (1999).

\bibitem{lambertson04}
G. Lambertson, private communication, 2004.

\end{thebibliography}
\end{document}